\documentstyle[sprocl,psfig]{article}

\def\vep{\varepsilon}
\def\ra{\rightarrow}

\def\be{\begin{equation}}
\def\ee{\end{equation}}
\def\bea{\begin{eqnarray}}
\def\eea{\end{eqnarray}}
\newcommand{\ba}{\begin{eqnarray}}
\newcommand{\ea}{\end{eqnarray}}

\def\vep{\varepsilon}

\makeatletter
\newbox\slashbox \setbox\slashbox=\hbox{\large$/$}
\def\pslash#1{\setbox\@tempboxa=\hbox{$#1$}
  \@tempdima=0.5\wd\slashbox \advance\@tempdima 0.5\wd\@tempboxa
  \copy\slashbox \kern-\@tempdima \box\@tempboxa}

\def\@cite#1#2{{#1\if@tempswa , #2\fi}}
\def\@citex[#1]#2{%
\if@filesw\immediate\write\@auxout{\string\citation{#2}}\fi
\leavevmode\unskip$^{\,\scriptstyle\@cite{\@collapse{#2}}{#1}}$}
\def\CITE{\@ifnextchar[{\@tempswatrue\@CITEX}{\@tempswafalse\@CITEX[]}}
\let\onlinecite\CITE
\def\@CITEX[#1]#2{%
\if@filesw\immediate\write\@auxout{\string\citation{#2}}\fi
\leavevmode\unskip\ \@cite{\@collapse{#2}}{#1}}
\def\@collapse#1{{%
\let\@temp\relax
\@tempcntb\@MM
\def\@citea{}%
\@for \@citeb:=#1\do{%
\@ifundefined{b@\@citeb}%
{\@temp\@citea{\bf ?}%
\@tempcntb\@MM\let\@temp\relax
\@warning{Citation `\@citeb ' on page \thepage\space undefined}}%
{\@tempcnta\@tempcntb \advance\@tempcnta\@ne
\edef\MyTemp{\csname b@\@citeb\endcsname}%
\def\@tempa{\@temptokena=\bgroup}%
\if0A{\fi%
\afterassignment\@tempa %
\@tempcntb=0\MyTemp\relax}%
\ifnum\@tempcntb=0\relax%
\@tempcntb=\@MM
\@citea\MyTemp
\let\@temp = \relax
\else %
\edef\@tempd{\number\@tempcntb}%
\ifnum\@tempcnta=\@tempcntb %
\ifx\@temp\relax %
\edef\@temp{\@citea\@tempd}%
\else
\edef\@temp{\hbox{--}\@tempd}%
\fi
\else %
\@temp\@citea\@tempd
\let\@temp\relax
\fi
\fi
}%
\def\@citea{, }%
}%
\@temp %
}}%

\makeatother
\begin{document}

\title{MONOPOLES AND CHAOS} 
\author{Harald Markum, Rainer Pullirsch, and Wolfgang Sakuler}
\address{Atominstitut, Technische Universit\"at Wien, A-1040 Vienna, Austria}
\maketitle\abstracts{ 
        We decompose U(1) gauge fields into a monopole and photon part
        across the phase transition from the confinement to the Coulomb
        phase.  We analyze the leading Lyapunov exponents of such
  	gauge field configurations on the lattice
        which are initialized by quantum Monte Carlo simulations.
        It turns out that there is a strong relation between the sizes
        of the monopole density and the Lyapunov exponent.
}

\sloppy
 
\section{Motivation}

The main question in QCD is the mechanism of confining quarks into
hadrons. One seeks for special classes of gauge fields which are responsible
for confinement. In the framework of the dual superconductor, color
magnetic monopoles arise as the appropriate candidates. Their investigation
necessitates a projection of the non-Abelian gauge theory on the compact
U(1) theory. In lattice computations it was demonstrated that monopoles
account for more than 90 percent of the string tension in the confinement
and they drop toward zero across the phase transition.

The study of chaotic dynamics of classical field configurations in field
theory finds its motivation in phenomenological applications as well 
as for the understanding of basic principles. 
The role of chaotic field dynamics for the confinement of quarks is a 
longstanding question. Here, we analyze the leading Lyapunov exponents 
of compact U(1) configurations on the 
lattice. The real-time evolution of the classical field equations 
was initialized from Euclidean equilibrium configurations created 
by quantum Monte Carlo simulations. This way we expect to see
a relationship between the appearance of monopoles and the strength of
chaotic behavior in lattice simulations.

\section {Monopoles in compact quantum field theories}

\begin{figure}[ht]
    \centerline{\psfig{figure=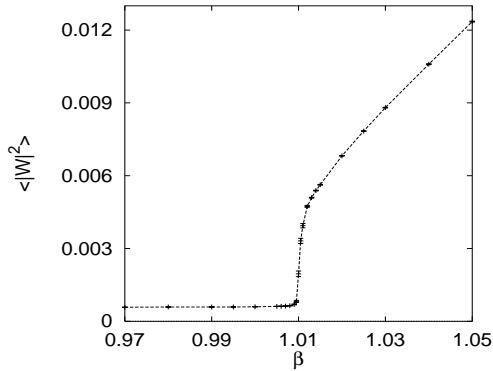,width=6cm,height=5cm}}
\vspace*{-2mm}
  \caption{Average of the absolute Polyakov loop squared $\langle | W |^2
           \rangle$ as a function of $\beta$ on a $12^4$ lattice.
           }
 \label{poly_abs_u1}
\end{figure}

We have investigated $4d$ U(1) gauge theory described by the action
\begin{equation}
S \lbrace U_l \rbrace = \sum_p (1 - \cos \theta_p ) \: ,
\label{decompeq}
\end{equation}
with $U_l = U_{x,\mu} = \exp (i\theta_{x,\mu}) $ and
$
  \theta_p =
 \theta_{x,\mu} +
 \theta_{x+\hat{\mu},\nu} -
 \theta_{x+\hat{\nu},\mu} -
 \theta_{x,\nu}\ \ (\nu \ne \mu)\ . $
At critical coupling 
$\beta_c \approx 1.01$ U(1) gauge theory undergoes a phase
transition between a confinement phase with mass gap and monopole
excitations for $\beta < \beta_c$ and the Coulomb phase 
with a massless photon for $\beta > \beta_c$,
see Fig.~\ref{poly_abs_u1}.

We are interested in the relationship between monopoles of
U(1) gauge theory and classical chaos across this phase transition. 
Following Refs.~\onlinecite{StWe92,Suzu96,Biel97,PLB01}, we have
factorized our gauge configurations
into monopole and photon fields.
The U(1) plaquette angles $\theta_{x,\mu\nu}$ are decomposed into the
``physical'' electromagnetic flux through the plaquette
$\bar \theta_{x,\mu\nu}$ and a number $m_{x,\mu\nu}$ of Dirac strings
passing through the plaquette
\begin{equation} \label{Dirac_string_def}
 \theta_{x,\mu\nu} = \bar \theta_{x,\mu\nu} + 2\pi\,m_{x,\mu\nu}\ ,
\end{equation}
where $\bar \theta_{x,\mu\nu}\in (-\pi,+\pi]$ and
$m_{x,\mu\nu} \ne 0$ is called a Dirac plaquette. Monopole and photon
fields are then defined in the following way
\begin{equation} \label{monoplaq}
\theta^{\rm mon}_{x, \mu} = - 2 \pi\, \sum_{x'} G_{x,x'} \,
\partial_{\nu}' \, m_{x', \nu\mu}
\end{equation}
\begin{equation} \label{photplaq}
\theta^{\rm phot}_{x, \mu} = - \, \sum_{x'} G_{x,x'} \,
\partial_{\nu}' \, \bar\theta_{x', \nu\mu} \ .
\end{equation}
Here $\partial'$ acts on $x'$, the quantities
$m_{x,\mu\nu}$ and $\bar\theta_{x, \mu\nu}$ are defined in
Eq.~(\ref{Dirac_string_def}) and $G_{x,x'}$ is the lattice Coulomb
propagator. One can show that $\tilde\theta_{x,\mu} \equiv
\theta^{\rm mon}_{x,\mu} +\theta^{\rm phot}_{x,\mu}$ is up to a gauge
transformation identical
with the original $\theta_{x,\mu}$ defined by $U_{x,\mu}=
\exp(i \theta_{x,\mu})$.

\section{Classical chaotic dynamics from quantum Monte Carlo
         initial states}

Chaotic dynamics in general is characterized by the
spectrum of Lyapunov exponents. These exponents, if they are positive,
reflect an exponential divergence of initially adjacent configurations.
In case of symmetries inherent in the Hamiltonian of the system
there are corresponding zero values of these exponents. Finally
negative exponents belong to irrelevant directions in the phase
space: perturbation components in these directions die out
exponentially. Pure gauge fields on the lattice show a characteristic
Lyapunov spectrum consisting of one third of each kind of
exponents.\cite{BOOK}
Assuming this general structure of the Lyapunov spectrum we
investigate presently its magnitude only, namely the maximal
value of the Lyapunov exponent, $L_{{\rm max}}$.

The general definition of the Lyapunov exponent is based on a
distance measure $d(t)$ in phase space,
\be
L := \lim_{t\ra\infty} \lim_{d(0)\ra 0}
\frac{1}{t} \ln \frac{d(t)}{d(0)}.
\ee
In case of conservative dynamics the sum of all Lyapunov exponents
is zero according to Liouville's theorem, $\sum L_i = 0$.
We utilize the gauge invariant distance measure consisting of
the local differences of energy densities between two $3d$ field configurations
on the lattice:
\be
d : = \frac{1}{N_P} \sum_P\nolimits \, \left| {\rm tr} U_P - {\rm tr} U'_P \right|.
\ee
Here the symbol $\sum_P$ stands for the sum over all $N_P$ plaquettes,
so this distance is bound in the interval $(0,2N)$ for the group
SU(N). $U_P$ and $U'_P$ are the familiar plaquette variables, constructed from
the basic link variables $U_{x,i}$,
\be
U_{x,i} = \exp \left( aA_{x,i}^cT^c \right)\: ,
\ee
located on lattice links pointing from the position $x=(x_1,x_2,x_3)$ to
$x+ae_i$. The generators of the group are
$T^c = -ig\tau^c/2$ with $\tau^c$ being the Pauli matrices
in case of SU(2) and $A_{x,i}^c$ is the vector potential.
The elementary plaquette variable is constructed for a plaquette with a
corner at $x$ and lying in the $ij$-plane as
$U_{x,ij} = U_{x,i} U_{x+i,j} U^{\dag}_{x+j,i} U^{\dag}_{x,j}$.
It is related to the magnetic field strength $B_{x,k}^c$:
\be
U_{x,ij} = \exp \left( \vep_{ijk} a B_{x,k}^c T^c \right).
\ee
The electric field strength $E_{x,i}^c$ is related to the canonically conjugate
momentum $P_{x,i} = \dot{U}_{x,i}$ via
\be
E^c_{x,i} = \frac{2a}{g^3} {\rm tr} \left( T^c \dot{U}_{x,i} U^{\dag}_{x,i} \right).
\ee

The Hamiltonian of the lattice gauge field system can be casted into
the form
\be
H = \sum \left[ \frac{1}{2} \langle P, P \rangle \, + \,
 1 - \frac{1}{4} \langle U, V \rangle \right].
\ee
Here the scalar product stands for
$\langle A, B \rangle = \frac{1}{2} {\rm tr} (A B^{\dag} )$.
The staple variable $V$ is a sum of triple products of elementary
link variables closing a plaquette with the chosen link $U$.
This way the Hamiltonian is formally written as a sum over link
contributions and $V$ plays the role of the classical force
acting on the link variable $U$. 

Initial conditions chosen randomly with a given average magnetic energy
per plaquette have been investigated in past years.\cite{ALGO.IJMP}
In the present study we prepare the initial field configurations
from a standard four dimensional Euclidean Monte Carlo program on
a $12^3\times 4$ lattice varying the inverse gauge coupling $\beta$.\cite{SU2}
We relate such four dimensional Euclidean
lattice field configurations to Minkow\-skian momenta and fields
for the three dimensional Hamiltonian simulation
by identifying a fixed time slice of the four dimensional lattice.

\section{Chaos and confinement  }

\begin{figure}[t]
\centerline{{\psfig{figure=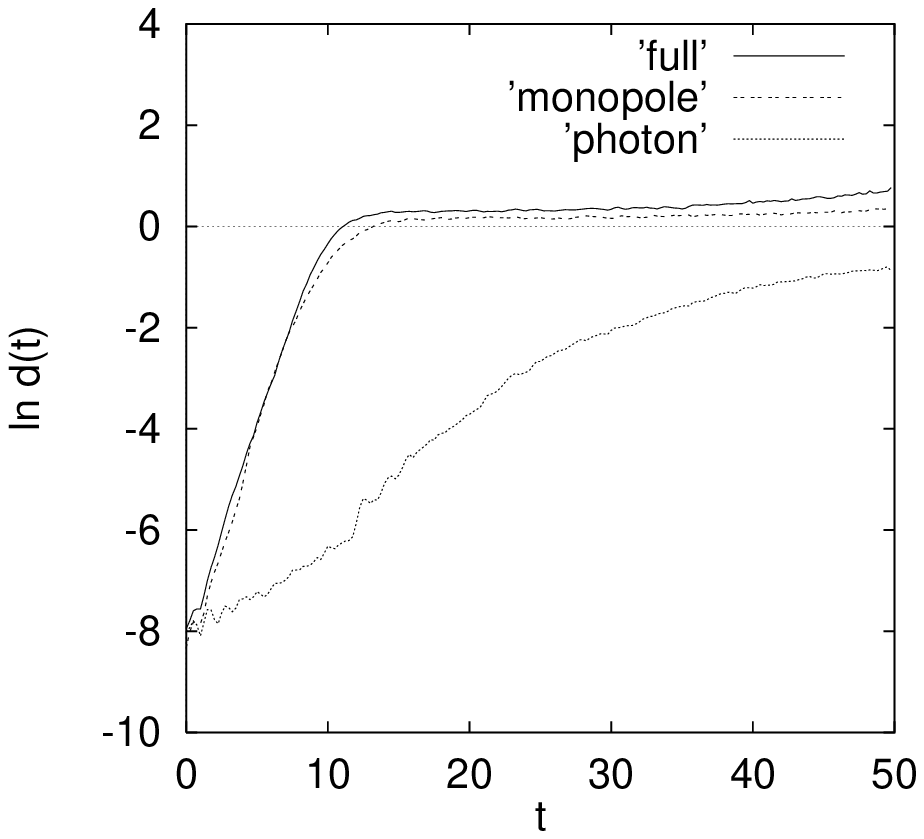,width=5cm}}\hspace{3mm}
{\psfig{figure=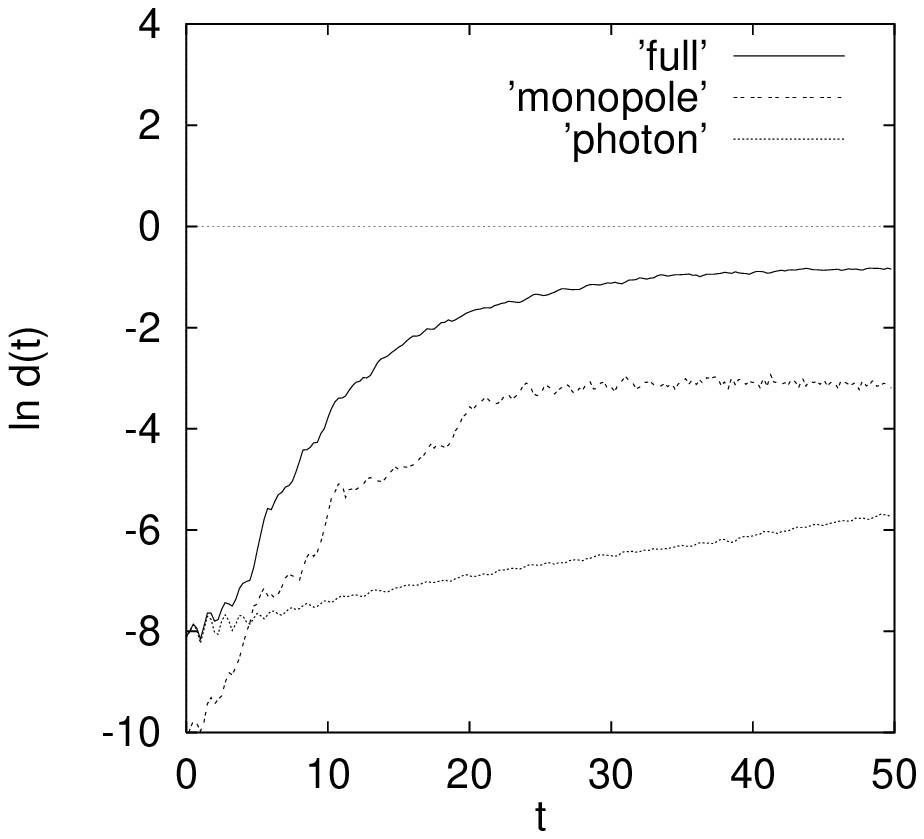,width=5cm}}}
\vspace*{-2mm}
\caption{
  Exponentially diverging distance of initially adjacent U(1) field
  configurations on a $12^3$ lattice prepared at $\beta=0.9$ in the
  confinement phase (left) and at $\beta=1.1$ in the Coulomb
  phase (right).
\label{Fig2}
 }
\end{figure}

We start the presentation of our results with a characteristic example
of the time evolution of the distance between initially adjacent
configurations. An initial state prepared by a standard four dimensional
Monte Carlo simulation is evolved according to the classical Hamiltonian dynamics
in real time. Afterwards this initial state is rotated locally by
group elements which are chosen randomly near to the unity.
The time evolution of this slightly rotated configuration is then
pursued and finally the distance between these two evolutions
is calculated at the corresponding times.
A typical exponential rise of this distance followed by a saturation
can be inspected in Fig.~\ref{Fig2} from an example of U(1) gauge theory
in the confinement phase and in the Coulomb phase.
While the saturation is an artifact of
the compact distance measure of the lattice, the exponential rise
(the linear rise of the logarithm)
can be used for the determination of the leading Lyapunov exponent.
The left plot exhibits that in the confinement phase the original
field and its monopole part have similar Lyapunov exponents while
the photon part has a smaller $L_{max}$. The right plot in the Coulomb
phase suggests that all slopes and consequently the Lyapunov
exponents of all fields decrease substantially.

\begin{figure}[t]
\centerline{{\psfig{figure=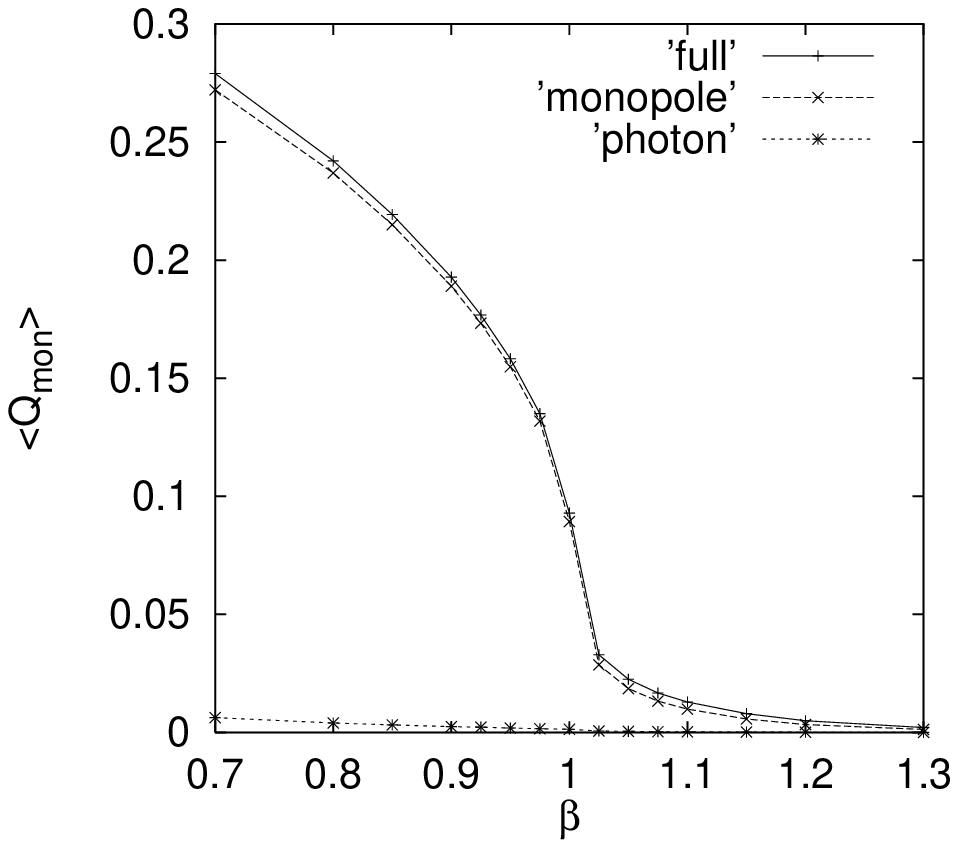,width=5cm}}\hspace{3mm}
{\psfig{figure=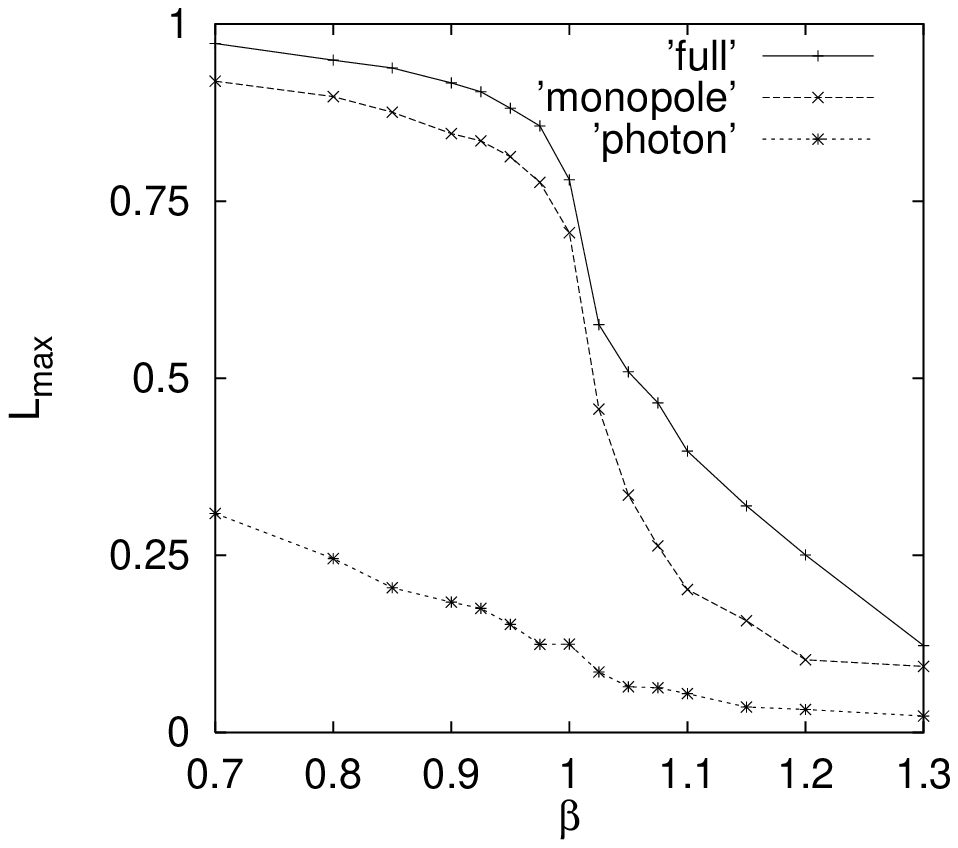,width=5cm}}}
\vspace*{-2mm}
\caption{Monopole density (left) and Lyapunov exponent
         (right) of the decomposed U(1) fields as a function
         of coupling.
\label{Fig3}
 }
\end{figure}

We now turn to a comparison with the monopole density in U(1)
quantum field configurations,
\be
Q_{mon}=\frac{1}{4V_4}\sum_{x,\mu\nu} |m_{x,\mu\nu}| \: .
\ee
The left plot of Fig.~\ref{Fig3} exhibits for a statistics of 100 
independent configurations that $Q_{mon}$ decreases
sharply  between the strong and the weak coupling regime.
It can be seen that the photon fields carry only a few monopoles.

The main result of the present study is the dependence of the leading
Lyapunov exponent $L_{{\rm max}}$ on the inverse coupling strength $\beta$,
displayed in the right plot of Fig.~\ref{Fig3}.
As expected the strong coupling phase, where confinement
of static sources has been established many years ago by proving the area law
behavior for large Wilson loops, is more chaotic.  The transition reflects
the critical coupling to the Coulomb phase. 
The right plot of Fig.~\ref{Fig3} shows that the monopole fields carry Lyapunov
exponents of nearly the same size as the full U(1) fields. The photon
fields yield a non-vanishing value in the confinement ascending toward $\beta=0$
for randomized fields which indicates that the decomposition (\ref{decompeq})
works perfectly for ideal configurations only.

\section{Conclusion}

We investigated the classical chaotic dynamics of
U(1) lattice gauge field configurations prepared by
quantum Monte Carlo simulation. The fields were decomposed
into a photon and monopole part. The maximal Lyapunov exponent
shows a pronounced transition as a function of the coupling strength
indicating that on a finite lattice configurations in the strong coupling
phase are substantially more chaotic than in the weak coupling regime.
The computations give evidence that the Lyapunov exponents in
the original U(1) field and in its monopole part are very similar.
The situation for the monopole density is analogous and serves as a 
consistency check of the decomposition. 
We conclude that classical chaos in field configurations and the
existence of monopoles are intrinsically connected to the confinement
of a theory.

\section*{ Acknowledgments  }
This work has been supported by the Austrian National Scientific Fund under
the project FWF P14435-TPH.
We thank Bernd A. Berg and Urs M. Heller as well as
Tam\'as S. Bir\'o and Natascha H\"ormann for previous cooperation concerning
topological objects and classical chaos in U(1) theory, respectively.

\end{document}